\documentstyle[prl,aps,epsfig]{revtex}

\begin{document}

\title{Measurement of the Beam-Spin Azimuthal Asymmetry\\
Associated with Deeply-Virtual Compton Scattering }

\vspace{4 mm}

\author{
A.~Airapetian,$^{31}$
N.~Akopov,$^{31}$
Z.~Akopov,$^{31}$
M.~Amarian,$^{26,31}$
E.C.~Aschenauer,$^{7}$
H.~Avakian,$^{11}$
R.~Avakian,$^{31}$
A.~Avetissian,$^{31}$
E.~Avetissian,$^{31}$
P.~Bailey,$^{15}$
% ...
\medskip\\ \centerline{(The HERMES Collaboration)}\medskip
}

\address{ 
% ...
$^7$DESY Zeuthen, 15738 Zeuthen, Germany\\
% ...
$^{11}$Istituto Nazionale di Fisica Nucleare, Laboratori Nazionali di Frascati, 00044 Frascati, Italy\\
% ...
$^{15}$Department of Physics, University of Illinois, Urbana, Illinois 61801, USA\\
% ...
$^{26}$Istituto Nazionale di Fisica Nucleare, Sezione Roma 1, Gruppo Sanit\`a and Physics Laboratory, Istituto Superiore di Sanit\`a, 00161 Roma, Italy\\
% ...
$^{31}$Yerevan Physics Institute, 375036, Yerevan, Armenia\\
} 

\vspace{4 mm}

\date{\today}

\maketitle

\begin{abstract}
The beam-spin asymmetry in hard electroproduction of 
photons has been measured for the first time. The data have been
accumulated by the HERMES experiment at DESY using the HERA 27.6 GeV
longitudinally polarized positron beam and an unpolarized hydrogen gas target.
The asymmetry in the azimuthal distribution of the produced photons 
in the angle $\phi$ relative to the lepton scattering plane was 
determined with respect
to the helicity state of the incoming positron beam. The beam-spin 
analyzing power in the $\sin\phi$ moment was measured to 
be -0.23 $\pm$ 0.04(stat) $\pm$ 0.03(syst) in the
missing-mass range below 1.7 GeV. 
The observed asymmetry is attributed to the interference of
the Bethe-Heitler and deeply-virtual Compton scattering processes.
\end{abstract}

\medskip

\vspace{4mm}

\centerline{PACS numbers: 13.60.Hb, 13.60.Le, 13.60.-r, 24.85.+p}

\twocolumn

The internal structure of the nucleon has been extensively studied
in deep-inelastic lepton scattering, resulting in such measurements
as the momentum distributions of quarks and their helicity 
dependences.
The contribution of the quark spins
to the nucleon spin was found to be small. 
Recently a possibility was identified to study
experimentally the total contributions of partons
to the nucleon spin, including their orbital angular momenta~\cite{bib:ji0}.
This idea is based on the formalism of the
so-called skewed parton distributions (SPD)
(also referred to as off-forward or generalized parton distributions
in the literature~\cite{bib:dittes,bib:muller,bib:rady1,bib:ji}).
In this formalism dynamical correlations between partons 
with different momenta are taken into account.
The SPD framework embodies a wide
range of observables, such as electromagnetic form factors,
conventional parton distributions and hard exclusive cross sections.
In particular, sum rules~\cite{bib:ji,bib:filip,bib:jaffe} relate 
second moments of certain SPDs with the total angular 
momenta of the quarks and of the gluons in the nucleon.

\begin{figure} [th]
\begin{center}
\epsfig{file=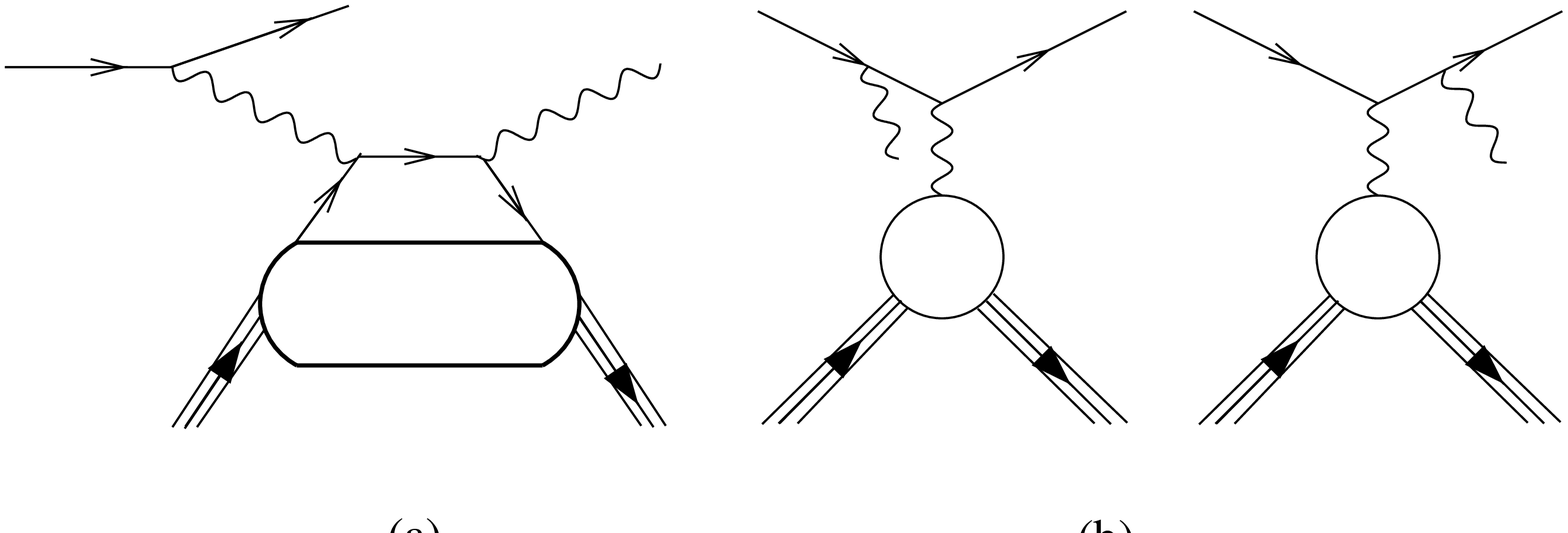,width=8.0cm}
\end{center}
\vspace{0.4cm}
\caption{(a) Feynman diagram for deeply-virtual Compton scattering, and
(b) photon radiation from the incident and scattered lepton in the
Bethe-Heitler process.}
\label{fig1}
\end{figure}

A reaction that can be cleanly interpreted
in terms of SPDs
is deeply-virtual Compton scattering (DVCS),
i.e. the exclusive leptoproduction
of a single multi-GeV photon with the target nucleon remaining intact.
Unfortunately, experimental information on DVCS is scant.
A central issue is that it is impossible
to distinguish between photons originating from DVCS and
those from the Bethe-Heitler (BH) process, which can be much more copious.
The corresponding diagrams 
are shown in Fig.~\ref{fig1}.
However, the interference between the DVCS and BH processes can
be exploited in order to obtain information on DVCS amplitudes.
For that purpose the HERMES collaboration has measured the
beam-spin asymmetry in hard exclusive electroproduction of photons.
The data obtained are presented in  this paper.

Using the notation
of Ref.~\cite{bib:diehl}, the cross section for exclusive leptoproduction of 
photons can be written as
\begin{eqnarray}
{\frac{d^4\sigma}{d \phi dt d{Q^2} d{x}}} =  {
        {\frac{x y^2}{32 \left(2 \pi\right)^4 Q^4}}
        {\frac{|{\tau^{}_{\mathrm{BH}}} + {\tau^{}_{\mathrm{DVCS}}}|^2}
              {{\left(1 + 4{x^2}{m^2}/{Q^2}\right)}^{1/2}}}
                                                },
\end{eqnarray}
\noindent
where $x$ represents the Bjorken scaling variable,
$y = \nu/E$ the fraction of the incident lepton energy $E$ carried by 
the virtual photon, $\nu$ its energy and $-Q^2$ its four-momentum squared, 
$m$ the proton mass, and $\tau_{\mathrm{BH}}$ 
and $\tau_{\mathrm{DVCS}}$ are the BH and DVCS amplitudes.
The cross section shown is differential in $x$, $Q^2$, $\phi$ and $t$,
where the azimuthal angle $\phi$ is the angle
between the lepton scattering plane and the plane defined by 
the virtual and real photons, and $t$ represents the square of
the four-momentum transfer to the target. 

In Ref.~\cite{bib:diehl} expressions are given 
for the DVCS+BH cross sections
in leading order ${\cal O}(1/Q)$. (An alternative approach
can be found in Ref.~\cite{bib:belit}, for instance.)
The leading-order interference term that 
depends on the helicity of the incident lepton is

\begin{eqnarray}
\nonumber
{({\tau^{*}_{\mathrm{BH}}}{\tau^{}_{\mathrm{DVCS}}} +
  {\tau^{*}_{\mathrm{DVCS}}}{\tau^{}_{\mathrm{BH}}})}_{pol} = 
        { { \frac{4\sqrt{2}~m e^6}{t Q x}  }
            \cdot { \frac{1}{\sqrt{1 - x}} } }\\
\times e_{\it l}P_{\it l} \Bigg[
 {-\sin \phi \cdot \sqrt{\frac{1+\epsilon}{\epsilon}} {\rm Im}\tilde{M}^{1,1}}
 \Bigg]. 
\end{eqnarray}
\noindent
The quantity $\tilde M^{1,1}$ 
is the linear combination of DVCS helicity amplitudes that
contributes in the case of a polarized beam and an unpolarized
target.
The interference is seen to depend on the azimuthal angle $\phi$,
the sign of the lepton charge $e_{\it l}$, and the polarization $P_{\it l}$
of the incident lepton. The kinematic quantity $\epsilon$ is
the polarization parameter of the virtual photon.
A determination of the $\sin\phi$-moment of the asymmetry
of the interference term shown in Eq.~(2) with respect to 
the beam polarization provides
information on the imaginary part of the DVCS amplitude combination
$\tilde{M}^{1,1}$, which is related to the SPDs~\cite{bib:diehl}.
Not shown in Eq.~(2) are other interference terms that are suppressed by 
${\cal O}(1/Q)$, but they involve other $\phi$-moments.

The data presented here were recorded during the 1996 and 1997 
running periods of the HERMES experiment using the 27.6 GeV HERA
longitudinally polarized positron beam at DESY~\cite{bib:barber}. 
The beam polarization was continuously measured by
Compton back scattering and had an average value of $0.55$ with 
a fractional uncertainty of 3.8\% \cite{bib:g1p,bib:longtrans}.  
The positrons were scattered off a hydrogen gas
target~\cite{bib:target}. Both unpolarized and spin-averaged 
polarized-target data have been used in the analysis. 

The scattered positrons and coincident photons were detected
by the HERMES spectrometer \cite{bib:spectrometer} in the 
polar-angle range of 40 to 220 mrad. 
A positron trigger was formed from a coincidence between
three scintillator hodoscope planes and a lead-glass calorimeter.
The trigger required an energy of more than 3.5 GeV deposited in the
calorimeter.
Charged particle identification was based on information
from four detectors: a threshold $\check{\rm C}$erenkov counter,
a transition radiation detector, a preshower scintillator counter
and a lead-glass calorimeter. The particle identification
provides an average positron identification efficiency of
99\% with a hadron contamination of less than 1\%.
Photons are identified by the detection of energy deposition in the
calorimeter and preshower counter without an associated charged track.

\begin{figure} [th]
\begin{center}
\epsfig{file=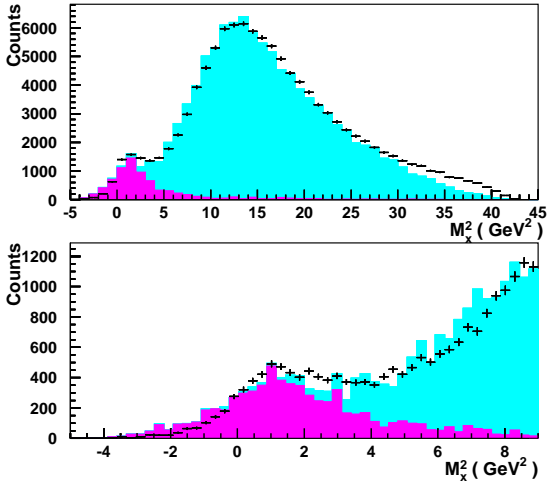,width=8.0cm}
\end{center}
\vspace{-1.5cm}
\caption{The measured distribution of photons observed in
hard electroproduction versus the missing mass squared ${M_x}^2$.
In the upper panel the full kinematic range is displayed,
while the low ${M_x}^2$ domain
is shown in the lower panel. The light-gray histogram represents
the results of a Monte-Carlo simulation in which fragmentation
processes and the Bethe-Heitler process are included, while
the dark-shaded histogram represents only the BH contribution.
The Monte-Carlo simulation includes the effect of the detector
resolution.}
\label{fig2}
\end{figure}

Events were selected if they contained only one positron track with
momentum larger than 3.5 GeV and only one photon with an energy 
deposition greater than 0.8 GeV in the calorimeter. The following
requirements were imposed on the positron kinematics: $Q^2$ $>$ 1 GeV$^2$,
$W^2$ $>$ 4 GeV$^2$, and $\nu$ $<$ 24 GeV, where $W$ denotes the
photon-nucleon invariant mass.

In Fig.~\ref{fig2},
the missing mass distribution of the selected events is
compared to the results of a Monte-Carlo (MC) simulation 
in which photons from fragmentation processes in
deep-inelastic scattering and from the exclusive BH process,
$e + p \rightarrow e' + \gamma + p$, are included.
The missing mass is defined as $M_x^2 = (q + P_p - k)^2$ with $q$,
$P_p$ and $k$ being the four-momenta of the virtual photon, the
target nucleon and the produced real photon, respectively. 
Due to the finite momentum resolution of the spectrometer
$M_x^2$ may be negative, in which case we define
$M_x = - \sqrt{-M_x^2}$.

The MC calculation is normalized to the same number of
deep-inelastically scattered positrons as were observed
inclusively in the experiment (about 5.1 million DIS events), which
corresponds to an integrated luminosity of 104 pb$^{-1}$.
There is fairly good agreement between the data and
the MC results in the relevant kinematic range of the
photon spectrum. In the region of low missing mass, the
main contribution is due to the BH process, while the
smeared DIS contribution is almost negligible. 
The smearing of the data to negative $M_x^2$ values is 
well reproduced by the Monte-Carlo calculation, which
does not include the DVCS process. 
This result is consistent

\begin{figure} [th]
\begin{center}
\epsfig{file=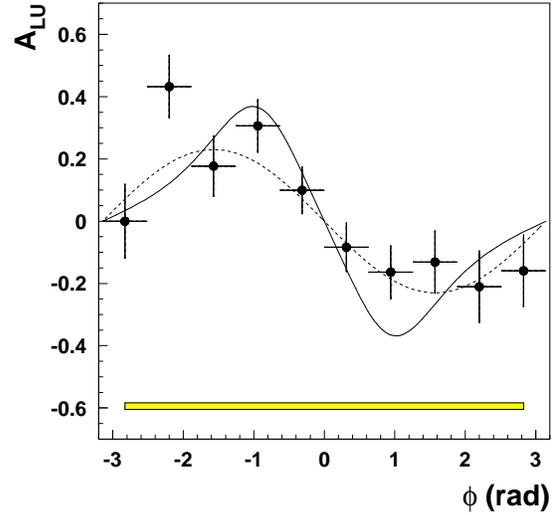,width=8.0cm}
\end{center}
\vspace{-1.2cm}
\caption{Beam-spin asymmetry $A_{LU}$ for hard electroproduction
of photons as a function of the azimuthal angle $\phi$. The data correspond
to the missing mass region between -1.5 and +1.7 GeV. The dashed curve
represents a $\sin\phi$ dependence with an amplitude of 0.23, while the
solid curve represents the result of a model calculation taken from
Ref.~[17]. The horizontal error bars represent the
bin width, and the error band below represents the systematic
uncertainty.}
\label{fig3}
\end{figure}

\noindent
with the calculations of Ref.~\cite{bib:vgg},
where it is shown that the DVCS contribution
to the electroproduction of photons is less 
than 10\% for the present kinematics.

Photon pairs from $\pi^0$ decay are removed from the data
by requiring the presence of exactly one photon cluster in the
(segmented) calorimeter. There could be a remaining
$\pi^0$ contamination from photon pairs that can not be
spatially resolved by the granularity of the calorimeter.
It may also happen that one of the $\pi^0$
decay photons escapes detection. These contaminations have been estimated
in the nominal exclusive region using a Monte-Carlo simulation.
It was found that $\pi^0$ mesons produced in exclusive processes
or as fragmentation products in deep-inelastic scattering may contaminate
the exclusive part of the photon spectrum by at most 8.5\%.
 
The DVCS-BH interference terms can be extracted
from the dependence of the data on the azimuthal angle $\phi$.
In order to have an almost full $\phi$-coverage,
events were selected with 15 $< \theta_{\gamma \gamma^{*}} <$ 70 mrad,
where $\theta_{\gamma \gamma^{*}}$ represents the angle
between the directions of the virtual photon and the real photon. 
A MC-simulation shows that for angles smaller than 15 mrad, the 
granularity of the calorimeter (9 $\times$ 9 cm$^2$)
is insufficient to reliably determine 
the angle $\phi$. For angles larger than 70 mrad, the $\phi$-acceptance 
is restricted. The average $\phi$-resolution in the selected
$\theta_{\gamma \gamma^{*}}$ range is about 0.14 rad.

In Fig.~\ref{fig3}, the azimuthal dependence of the
measured beam-spin asymmetry $A_{LU}$ is shown, which
is defined as
\begin{eqnarray}
A_{LU}(\phi)  = \frac{1}{\langle |P_l| \rangle} \cdot
		\frac{N^+(\phi) - N^-(\phi)} {N^+(\phi) + N^-(\phi))},
\end{eqnarray}

\begin{figure} [th]
\begin{center}
\epsfig{file=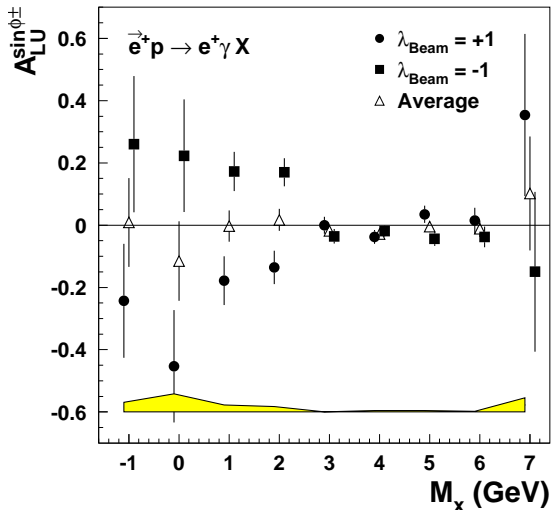,width=8.0cm}
\end{center}
\vspace{-1.5cm}
\caption{The $\sin\phi$-moment $A_{LU}^{\sin\phi^{\pm}}$
as a function of the missing mass for positive beam helicity (circles),
negative beam helicity (squares) and the averaged helicity (open
triangles). A negative value is assigned to $M_x$ if $M_x^2 <$ 0.
The error bars are statistical only. The systematic uncertainty is
represented by the error band at the bottom of the figure.}
\label{fig4}
\end{figure}

\noindent
where $N^+$ and $N^-$ represent the luminosity-normalized
yields of events 
with corresponding beam helicity states, $\langle |P_l| \rangle$
is the average magnitude of the beam polarization, and the
subscripts $L$ and $U$ denote a longitudinally
polarized beam and an unpolarized target.
The data displayed in Fig.~\ref{fig3} have been
selected requiring a missing mass between -1.5 and +1.7 GeV, i.e.
-3$\sigma$ below and +1$\sigma$ above $M_x = m$, and represent
4015 events. An asymmetric
$M_x$-range was chosen to minimize the influence of
the DIS-fragmentation background while optimizing the
statistics. Both the proton and the $\Delta(1232)$-resonance
are included in the selected $M_x$ range.
However, the data most likely originate from
the exclusive final state with one proton, since
the scattering process is dominated by the elastic
contribution at kinematics relevant for the BH process. 
This conclusion is supported experimentally
by figure 2, where the elastic BH Monte Carlo gives 
a good account of the photon spectrum at low $M_x$, and
theoretically in Ref.~\cite{bib:strikman}.
The comparison of the $A_{LU}$ data in Fig.~\ref{fig3} to
a simple $\sin\phi$ curve demonstrates that the data
have the $\phi$-dependence expected from Eq.~(2). 
The model calculation of Ref.~\cite{bib:kivel} which is
based on the SPD framework and computed at the average 
kinematics of the present experiment has also been displayed.

In order to be able to compare the $\phi$-dependence of 
the beam-spin asymmetry for various missing mass bins, 
the $\sin\phi$-weighted moments have been determined:
\begin{eqnarray}
	A_{LU}^{\sin\phi^{\pm}} = 
		\frac{2}{ N^{\pm} }
	 	\sum_{i=1}^{N^{\pm}} \frac{\sin\phi_i}{|P_l|_i},
\end{eqnarray}

\noindent
where the superscript $\pm$ refers to the helicity of the positron beam.
In Fig.~\ref{fig4} the extracted values of $A_{LU}^{\sin\phi^{\pm}}$

\begin{figure} [th]
\begin{center}
\epsfig{file=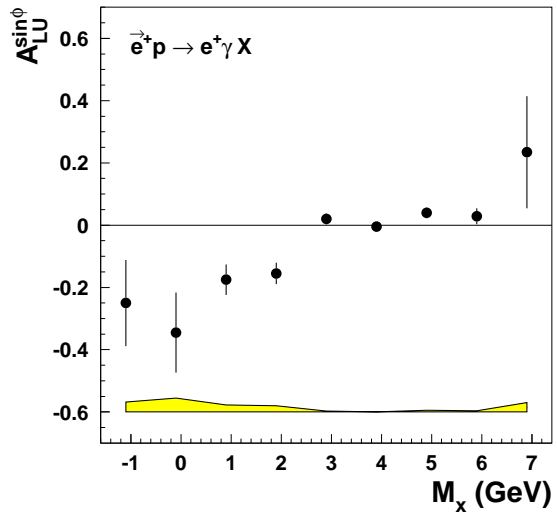,width=8.0cm}
\end{center}
\vspace{-1.5cm} 
\caption{The beam-spin analyzing power $A_{LU}^{\sin\phi}$ for
hard electroproduction of photons on hydrogen as a function of
the missing mass. The systematic uncertainty is represented by
the error band at the bottom of the figure.}
\label{fig5}
\end{figure}

\noindent
are plotted versus the missing mass $M_x$
for the two helicity states $\lambda_{Beam}$ of the positron beam.
The sign of the $\sin\phi$-moment
is opposite for the two beam helicities, in
agreement with the expectations for the helicity
dependence of the relevant DVCS-BH interference term.
The beam-spin averaged data are consistent with zero,
which 
is in agreement with the expectations for unpolarized beam
and target. The beam-spin averaged data can be used to determine
an upper limit of a possible false asymmetry due to instrumental
effects which --- averaged for
$M_x$ between -1.5 and +1.7 GeV --- amounts to -0.03 $\pm$ 0.04.

As the data in Fig.~\ref{fig4} for the two beam-helicity states
contain the same physics information, they are combined when
evaluating the
beam-spin analyzing power $A_{LU}^{\sin\phi}$:
\begin{eqnarray}
        A_{LU}^{\sin\phi} =
        \frac{2}{ N }
        \sum_{i=1}^{N} \frac{\sin\phi_i}{(P_l)_i},
\end{eqnarray}

\noindent
where $N = N^+ + N^-$. In contrast to Eq.~(4), the sign of the
beam polarization is explicitly taken into account, thus distinguishing the
two helicity states. The results are presented
in Fig.~\ref{fig5} versus missing mass. All bins
in the missing mass region below $M_x \approx$ 2.5 GeV show a
similar negative asymmetry,
while $A_{LU}^{\sin\phi}$ is consistent with zero for larger
$M_x$ values. Consequently smeared DIS events at low $M_x$ can only 
marginally dilute the observed asymmetry. As the missing-mass
resolution of the HERMES spectrometer for DVCS-like events
is $\sim$ 0.77 GeV, part of
the exclusive data falls below or above $m$.
As a result the missing-mass bins left and right of $M_x = m$
also show a non-zero value of $A_{LU}^{\sin\phi}$
in Fig.~\ref{fig5}. 

In order to evaluate the systematic uncertainty on $A_{LU}^{\sin\phi}$
several contributions were considered. The same
MC simulation described above 
has been used to estimate the smearing effect
on $A_{LU}^{\sin\phi}$, which was found to
be less than 5\%. The systematic uncertainty associated with
smearing and beam polarization is represented by
the error bands displayed in
Figs.~\ref{fig3},~\ref{fig4},~\ref{fig5}. In the exclusive
region (-1.5 $<$ $M_x$ $<$ 1.7 GeV), two additional contributions
to the systematic uncertainty were considered.
Possible false asymmetries due to the BH process
are at most 2.6\%, while the uncertainty due to the 
$\pi^0$ contamination is estimated to be 12.5\%.
The total systematic uncertainty
at $M_x \approx m$ amounts to 0.03.
The quoted instrumental false asymmetry 
has not been included in this number
as it cancels in $A_{LU}^{\sin\phi}$.

By combining
the $A_{LU}^{\sin\phi}$ data in the same $M_x$ region
as was used for Fig.~\ref{fig3}
(-1.5 $<$ $M_x$ $<$ 1.7 GeV), an average value
of -0.23 $\pm$ 0.04 (stat) $\pm$ 0.03 (syst) is obtained.
The average values of the kinematic variables corresponding
to this measurement are: $\langle x \rangle$ = 0.11,
$\langle Q^2 \rangle$ = 2.6 GeV$^2$ and $\langle -t \rangle$ 
= 0.27 GeV$^2$. Since the BH process is dominated by
the exclusive final state with one proton~\cite{bib:strikman},
and interference
can only occur between processes with identical final
states, the measured beam-spin analyzing power 
can be compared to calculations for
exclusive processes which are based on the SPD framework.
In Ref.~\cite{bib:kivel}, e.g.,
a value of -0.37 is quoted for $A_{LU}^{\sin\phi}$ in
a calculation for kinematics close to those of the present 
experiment. This calculation includes a twist-3 contribution of
less than 5\%.

In summary, the beam-spin azimuthal asymmetry for
hard electroproduction
of photons has been measured in the missing mass ($M_x$)
range up to 7 GeV. A non-zero
asymmetry is observed in the exclusive domain, i.e.
for $M_x \leq$ 1.7 GeV. The observed $\sin\phi$-moment of the data has
the beam-helicity dependence expected from interference 
between deeply-virtual Compton scattering and the Bethe-Heitler 
process. 

We gratefully acknowledge the DESY management for its support and
the staffs at DESY and the collaborating institutions for their
significant effort, and our funding agencies for financial support.


\begin{references}

\bibitem{bib:ji0}  X. Ji, Phys. Rev. Lett. {\bf 78} (1997) 610.

\bibitem{bib:dittes} F-M. Dittes {\it et al.}, Phys. Lett. 
{\bf B 209} (1988) 325. 

\bibitem{bib:muller} D. M\"uller {\it et al.}, Fortsch. Phys. 
{\bf 42} (1994) 101. 

\bibitem{bib:rady1} A.V. Radyushkin, Phys. Lett. {\bf B 385} (1996) 333.

\bibitem{bib:ji}  X. Ji, Phys. Rev. {\bf D 55} (1997) 7114.

\bibitem{bib:filip}  B.W. Filippone and X. Ji, Adv. in Nucl. Phys. (2001),
in press [hep-ph/0101224].

\bibitem{bib:jaffe} R.L. Jaffe, Nucl. Phys. {\bf B 536} (1998) 303.

\bibitem{bib:diehl} M. Diehl {\it et al.}, Phys. Lett. {\bf B 411} (1997) 193.

\bibitem{bib:belit} A.V. Belitsky {\it et al.}, Nucl. Phys. {\bf B 593} 
(2001) 289.

\bibitem{bib:barber} D.P. Barber {\it et al.}, Phys. Lett. {\bf B 343} 
(1995) 436.

\bibitem{bib:g1p} HERMES Collaboration, A. Airapetian {\it et al.},
Phys. Lett. {\bf B 442} (1998) 484.

\bibitem{bib:longtrans} M. Beckmann {\it et al.}, 
  Nucl. Instr. Meth. (2001), in press [physics/0009047].

\bibitem{bib:target} J. Stewart. Proc. of the Workshop "Polarized
gas targets and polarized beams" eds. R.J. Holt and M.A. Miller,
Urbana-Champaign, AIP Conf. Proc. {\bf 421} (1997) 69.

\bibitem{bib:spectrometer} HERMES Collaboration, K.Ackerstaff {\it et al.},
  Nucl. Instr. Meth. {\bf A 417} (1998) 230.

\bibitem{bib:vgg} P.A.M. Guichon and M. Vanderhaeghen, Prog. Nucl. Part.
  Phys. {\bf 41} (1998) 125; M. Vanderhaeghen, P.A.M. Guichon and M. Guidal,
  Phys. Rev. Lett. {\bf 80} (1998) 5064.

\bibitem{bib:strikman} L.L. Frankfurt, M.V. Polyakov, and M. Strikman, 
	Contribution to the Workshop "Jefferson Lab Physics and 
	Instrumentation with 6-12 GeV Beams", Newport News, VA, 
	June 15-18, 1998, hep-ph/9808449.

\bibitem{bib:kivel} N. Kivel, M. Polyakov and M. Vanderhaeghen,
	Phys. Rev. {\bf D 63} (2001) 114014.

\end{references}
\end{document}